\documentclass[final,5p,times,twocolumn,authoryear]{elsarticle}
\usepackage{natbib}
\setcitestyle{numbers,square}
\usepackage{lmodern}
\usepackage{soul}
\usepackage{amsmath,amssymb,amsfonts}
\usepackage{mathtools}
\usepackage{color}
\usepackage{diagbox}
\usepackage{textcomp}
\usepackage{array}
\usepackage{silence}
\usepackage{graphicx} 			
\usepackage{subcaption}			
\usepackage{amsfonts,amsmath,amssymb}
\usepackage[colorlinks, citecolor=blue]{hyperref}
\usepackage{url}            
\usepackage{booktabs}       
\usepackage{amsfonts}       
\usepackage{nicefrac}       
\usepackage{microtype}      
\usepackage{lipsum}
\usepackage{colortbl}
\usepackage{booktabs,tabulary,multirow,overpic,xcolor}
\usepackage{hyperref}
\usepackage{float}
\usepackage{fancyhdr}       
\usepackage{xcolor}
\usepackage{hyperref}
\usepackage{amsthm}
	
	\newtheorem{proposition}{Proposition}
	\newenvironment{Proof}{{\noindent\bf Proof:~}}{\hfill $\blacksquare$\par}

\usepackage{bm}
\usepackage{fancyhdr,amsmath,amssymb}
\usepackage[ruled,vlined,linesnumbered]{algorithm2e}

\usepackage{xifthen}
\def\BibTeX{{\rm B\kern-.05em{\sc i\kern-.025em b}\kern-.08em
    T\kern-.1667em\lower.7ex\hbox{E}\kern-.125emX}}
\AtBeginDocument{\definecolor{tmlcncolor}{cmyk}{0.93,0.59,0.15,0.02}\definecolor{NavyBlue}{RGB}{0,86,125}}

\definecolor{color3}{rgb}{0.95,0.95,0.95}
\newcommand{\figref}[1]{Fig. \ref{#1}}

\begin{document}

\begin{frontmatter}

\date{}

\title{Identification of Path Congestion Status for Network Performance Tomography using Deep Spatial-Temporal Learning}


\author[first]{Chengze Du \corref{cor1}}
\author[first]{Zhiwei Yu}
\author[first]{Xiangyu Wang}
\affiliation[first]{organization={Beijing University of Posts and Telecommunications (BUPT)} ,         
          city={Beijing},
          postcode={100876},
          country={China}}
          
\begin{abstract}
Network tomography plays a crucial role in assessing the operational status of internal links within networks through end-to-end path-level measurements, independently of cooperation from the network infrastructure. However, the accuracy of performance inference in internal network links heavily relies on comprehensive end-to-end path performance data. Most network tomography algorithms employ conventional threshold-based methods to identify congestion along paths, while these methods encounter limitations stemming from network complexities, resulting in inaccuracies such as misidentifying abnormal links and overlooking congestion attacks, thereby impeding algorithm performance.
This paper introduces the concept of Additive Congestion Status to address these challenges effectively. Using a framework that combines Adversarial Autoencoders (AAE) with Long Short-Term Memory (LSTM) networks, this approach robustly categorizes (as uncongested, single-congested, or multiple-congested) and quantifies (regarding the number of congested links) the Additive Congestion Status. Leveraging prior path information and capturing spatio-temporal characteristics of probing flows, this method significantly enhances the localization of congested links and the inference of link performance compared to conventional network tomography algorithms, as demonstrated through experimental evaluations.
\end{abstract}



\begin{keyword}
Network Tomography\sep Path Congestion Status\sep End-to-End Measurement\sep LSTM\sep Adversarial Autoencoders


\end{keyword}

\end{frontmatter}




\section{INTRODUCTION}

{A}{s} networks grow in size and complexity\cite{review_dl_abw_Sun_2023}, sophisticated and minimally invasive tools for network performance assessment become ever more necessary. Network tomography\cite{review_nt_2004} is an innovative approach that leverages end-to-end measurements(e.g., delays or packet loss rates) to infer the status of internal network links by using information from probing between source-end hosts. 

Measuring the information of an end-to-end path performance directly determines whether the performance status information of the internal links in the target network can be successfully inferred. When the end-to-end path performance measurement is under perfect error-free conditions, the performance parameters of the internal network links can theoretically be accurately inferred. Nevertheless, in real network environments, end-to-end measurements are influenced by various error factors, forcing the path performance parameters to deviate from their true values with a certain error degree. Therefore, improving the overall detection accuracy of end-to-end paths is necessary to ensure the reliability and effectiveness of inferring the performance parameters of network tomography links and to obtain more accurate path-performance parameter measurement data.

End-to-end path performance parameter measurements can be conducted either through active measurement or passive observation. Although the latter avoids introducing any detection load cost, its limitations in operability and controllability have promoted active measurement, which is currently more commonly used.

	

\begin{figure*}[t]
        \label{fig: intro}
	\centering
	\begin{minipage}[c]{0.55\textwidth}
            \centering
		\includegraphics[width=\textwidth]{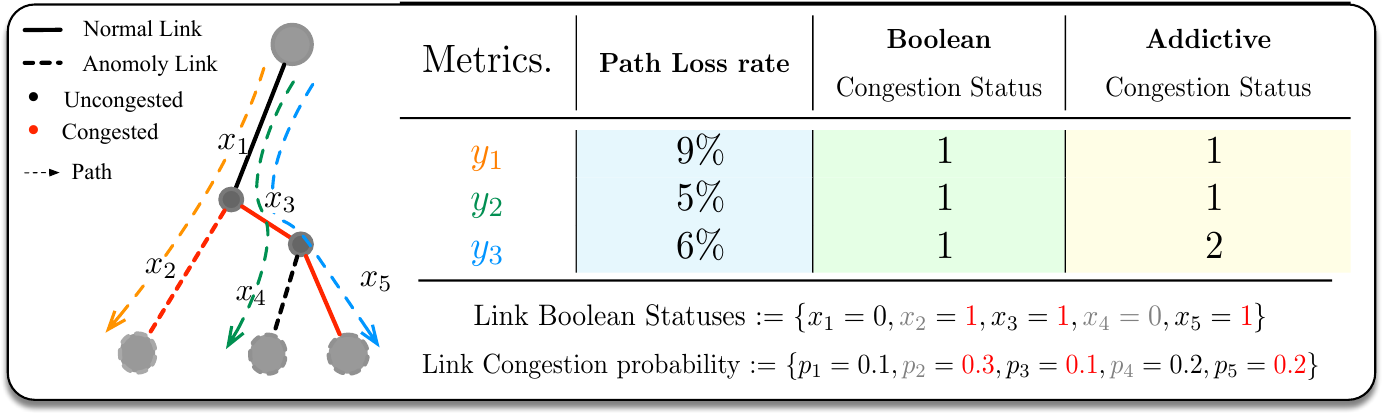}
		\subcaption{}
		\label{fig_E2_1}
	\end{minipage}
	\begin{minipage}[c]{0.42\textwidth}
            \centering
		\includegraphics[width=\textwidth]{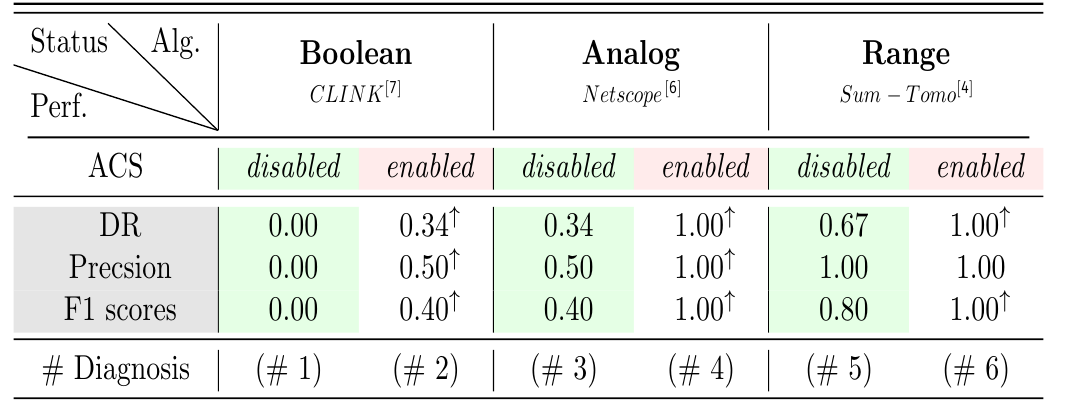}
		\subcaption{}
		\label{fig_E2_2}
	\end{minipage} \\
 	\centering
	\begin{minipage}[c]{\textwidth}
            \centering
		\includegraphics[width=\textwidth]{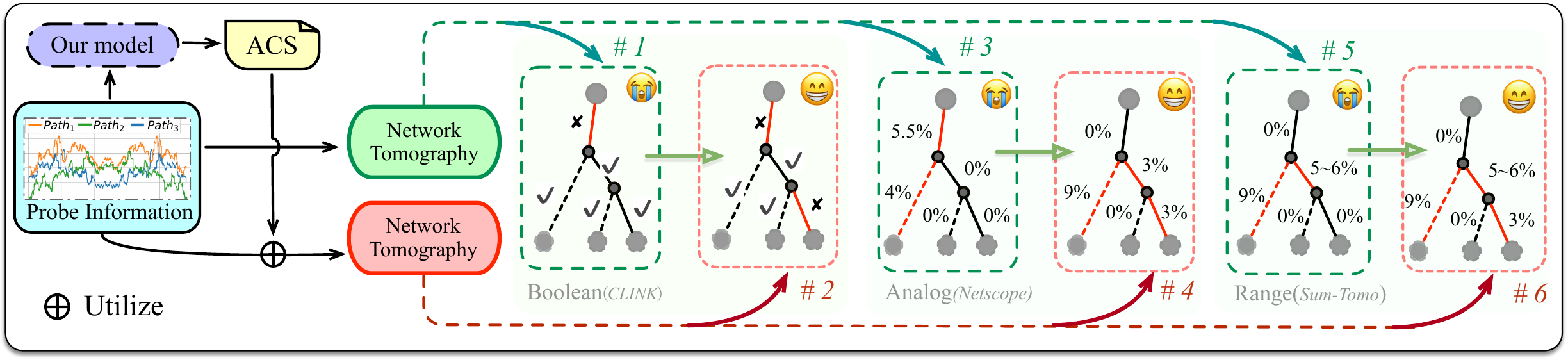}
		\subcaption{}
		\label{fig_E3_1}
	\end{minipage}
        \caption{
    Illustration of Addictive Congestion Status (ACS) based network boolean tomography for congested link identification. 
    (a) A network composed of four links and three paths. The observed status of each path is quantified by their loss rates.
    (b) Diagnosis performance comparisons of different network tomography schemes. 
    (c) Illustration of the enhancement when applying ACS for path status quantification compared to conventional boolean ones.}
\end{figure*}

Each method has a varying requirement for end-to-end path performance measurement information. Specifically, early network tomography relied on highly invasive probing to deduce specific properties of links, assuming that if two paths traverse a single, shared lossy link, their measured path loss rates would be equal. For example, network Boolean tomography \cite{NBT_Duffield_2006} employed lighter probing methods to infer the Boolean status of links based on rough path information. Such schemes rely on heuristic thresholds based on delay and packet loss performance to determine paths' binary congestion status (congested or none-congested). New frameworks, such as Range tomography \cite{range_nt}, utilize multiple detection data from end-to-end paths, e.g., packet loss rates and available bandwidth, to balance probing cost and accuracy.

Based on a simple observation in normal network scenarios, a congested link will cause all paths passing through it to become congested. Therefore, network Boolean tomography extracts the link status by reversing the binary path status. However, it is ill-posed, as multiple sets of congested links (solutions) often match the observed congestion on the end-to-end paths, limiting its diagnostic performance. However, it should be noted that once a path traverses multiple congested links, its packet loss rate and communication delay are much higher than those of a single congested link. Hence, we identify congestion by differentiating between the congestion status of the paths to reveal the varying numbers of congested links they encounter. Compared to binary congestion status, multi-dimensional congestion status reduce the size of the solution space, thereby improving diagnostic performance.

Based on this, this paper  distinguishes and identifies the congestion status of paths by proposing a metric called Addictive Congestion Status and a method to measure it. Qualitatively, the metric encompasses three status: none-congested, single-congested, and multi-congested. Quantitatively, it represents the number of congested links on the path. Furthermore, we develop a new observation framework integrating adversarial autoencoders \cite{makhzani2015adversarial} with Long Short-Term Memory (LSTM) networks to refine the network tomography process. This hybrid approach leverages the adversarial autoencoder's ability to learn complex, non-linear representations of network paths and the LSTM's proficiency in capturing temporal dependencies within the path measurements. This strategy accurately identifies the Addictive Congestion Status qualitatively and quantitatively. Finally, we validate the contribution of Addictive Congestion Status to existing methods by selecting three classic network tomography algorithms: Netscope\cite{ant-5-Ghita_Nguyen_Kurant_Argyraki_Thiran_2010}, CLINK\cite{Nguyen_Thiran_2007_clink}, and Sum-Tomo\cite{range_nt}, which respectively belong to Analog Tomography, Boolean Tomography, and Range Tomography. The results demonstrate a significant improvement in the accuracy of network tomography techniques regarding congested link localization and link performance inference.
The majore contributions of our work are summarized as follows: 

\begin{itemize}
        \item \textbf{Novel Hybrid Model} We introduce a new method that combines adversarial autoencoders (AAE) with  LSTM networks to redefine the Addictive Congestion Status (ACS) and improve network traffic classification and quantification by leveraging spatio-temporal characteristics.
    \item \textbf{Enhanced Network Tomography Performance} The proposed approach significantly improves network tomography, reducing false and missed congestion detections and improving metrics such as recall, precision, F1 score, and normalized root mean square error (NRMSE) in link performance inference.
\end{itemize}

\section{RELATED WORK}

Reliably assessing link performance within a network infrastructure is crucial for maintaining optimal service quality\cite{network-service-qos-wustrow2011telex}. For example, congested links can significantly impair network services, affecting data transmission rates and increasing latency. However, directly measuring the individual link performance imposes considerable challenges. The primary obstacles include the high costs\cite{nbt-vs-ant-Nguyen_Thiran_2005} associated with measurement processes and the administrative authority required to access certain parts of the network\cite{network-privacy-Donnet_Friedman_2007}. Furthermore, direct measurement approaches raise privacy concerns, necessitating deep visibility into network traffic patterns. Thus, network tomography has emerged as a pivotal methodology in light of these challenges.

Network tomography innovatively infers the performance status of internal links using aggregated end-to-end measurements\cite{review_nt_2004} and minimizing the need for invasive monitoring techniques. Analog network tomography uses invasive probing methods to obtain as many messages as possible to solve the under-constrained linear equations between links and paths\cite{ant-1-Caceres_Duffield_Moon_Towsley_2003}. Most methods employ additional optimization and constraints to enhance their performance \cite{ant-2-Chen_Cao_Bu_2007,ant-5-Ghita_Nguyen_Kurant_Argyraki_Thiran_2010, ant-3-Sossalla, ant-4-qiao}. It should be noted that analog methods impose high requirements in probing and computational resources \cite{NBT_Duffield_2006}. Besides, Boolean tomography adopts binary path measurements to deduce the link status ('Bad' or 'Normal')\cite{nbt-1-Duffield_2003,NBT_Duffield_2006,nbt-2-ogino, nbt-3-mukamoto, nbt-4-bartolini}. This method reduces the data volume and privacy intrusions at the expense of measurement precision\cite{nbt-vs-ant-Nguyen_Thiran_2005}. However, considering that the adaptability of different network services to different congestion levels is vastly different, some new network tomography frameworks, such as Range tomography, identify the boolean status and the congestion range\cite{range_nt}. Furthermore, some recent papers\cite{nbt-adv-1, nbt-adv-2, nbt-adv-3} focus on attack methods targeting network tomography monitoring. These works demonstrate that attackers can deliberately increase the forwarding delay on victim links after illicitly gaining control over some network nodes while manipulating the inferred results of network tomography. This allows them to conceal their attacks, which degrade link performance.

Long Short-Term Memory\cite{lstm} (LTSM) networks are a specialized form of recurrent neural networks\cite{rnn} uniquely designed to address the challenges associated with learning from data sequences \cite{review-lstm-karim2019insights}. Their architecture enables them to remember information for long periods, essential for capturing the temporal dependencies inherent in network data. Several existing works have demonstrated the appealing performance of LSTM in prediction and classification problems\cite{lstm-1-tejasree2024land, lstm-2-saurabh2024deep, lstm-3-dou2023memristor, lstm-4-srikantamurthy2023classification, lstm-5-sun2024automatic, lstm-6-li2024demand}, demonstrating a very low error rate in predicting epidemic diseases\cite{covid1-Arora_Kumar_Panigrahi_2020, covid2-Shahid_Zameer_Muneeb_2020, covid3-Chimmula_Zhang_2020}. Subsequently, LSTM has been integrated with other modules. For instance, Firzt\cite{Fritz_Dorigatti_Rügamer_2021} combined GNN and LSTM to predict COVID-19 cases in Germany. Chen\cite{Chen_Xu_Wu_Zheng_2018} proposed GC-LSTM, a Graph Convolution Network (GC) embedded LTSM for end-to-end dynamic link prediction. Li \cite{lstm-6-li2024demand} combined convolutional neural networks (CNN) and long short-term memory to predict the Nev sales and charge infrastructure ownership, which shows the best performance in using computing power.  Sun \cite{lstm-5-sun2024automatic} introduced a deep learning model based on LSTM to improve the shortcomings of existing automatic driving trajectory planning systems, which enhances both the accuracy and safety of the output trajectories. Alternatively, adversarial-LSTM and LSTM-adversarial autoencoders use LSTM to train the latent space produced by autoencoder\cite{adv-lstm-Casas_Arcucci_Mottet_Guo_Pain_2021, lstm-adv-Ouyang_Wang_Zhao_Wu_2021}, and our work is different from them.

\section{PRELIMINARIES}

\subsection{Network Model} 
Network topology is considered an undirected graph $ \mathcal{G}=(\mathcal{N},\mathcal{L})$, where nodes $\mathcal{N}$ can be routers, switches, and hosts. The source node is typically the host that sends the data flow, while the edge node usually receives the data flow. The direct connection between nodes is called a link $l$, and the set of all links in the topology is $\mathcal{L}$. A data flow from the sending host to the receiving host often passes through multiple nodes, traversing several unknown links. Besides, a path $p$ is an ordered set of links between a pair of source and edge nodes, and the set of all paths in the topology is $\mathcal{P}$.

To better express the relationship between paths and links, a routing matrix R is employed to elucidate. The routing matrix is a two-dimensional Boolean array of size $|\mathcal{P}| \times |\mathcal{L}|$, where the row coordinates represent path numbers, and the column coordinates are link numbers. If the number in the i-th row and j-th column of R is 1, the i-th path in the network topology includes the j-th link. The set of row coordinates with the number 1 in the n-th column vector represents the set of link coordinates that the n-th path routes through.

Before clarifying the Path Congestion Statues, it is necessary to state the following two assumptions:
\begin{enumerate}
  \item R does not change during the measurement process, and observations that do not match R are discarded.
  \item The states of the links are independent of each other and do not affect each other.
  \item The network status remains stable and does not change drastically over a period of time.
\end{enumerate}

\subsection{Probing Flow} 
The probing flow mechanism was initially used to measure a path's available bandwidth by identifying it based on the characteristics of the probing flow. It is categorized into a GAP model and RATE model based on the construction of the probing flow\cite{probe-model}, which is achieved by increasing the probing flow size to induce self-congestion while considering the probing flow size at the moment of self-congestion as the available bandwidth. This approach often exacerbates the congestion situation on links. 

\begin{figure}[t]
	\centering
	\includegraphics[width=\columnwidth]{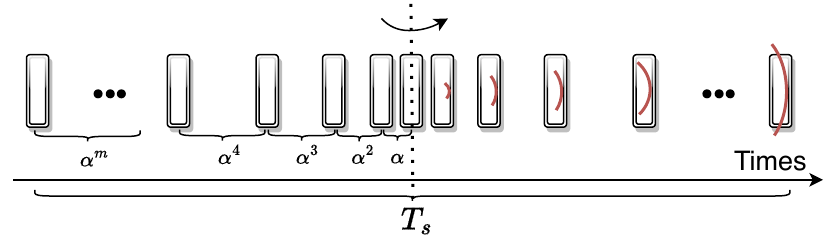}
	\caption{Illustration of the probing flow.}
	\label{fig: Probeflow}
\end{figure}

The probing flow features include packet delay size, variance of all delays, changes in packet gaps, and packet loss rate. This work aims to capture the congestion status of a path through these characteristics, namely, how many links along the path are congested.
\figref{fig: Probeflow} highlights that the developed framework adopts an exponential packet-sending mechanism where the interval between each packet sent is $\alpha^m, \alpha^{m-1}, \alpha^2, \alpha$. Compared to constant intervals, exponential intervals use the same number of packets but are less intrusive\cite{probe-Feat-Wang_Cheng_2006}. Thus, we present $\alpha$ to limit the average probing flow rate to 1\%. For experimental verification, refer to Section 5.

For each probe packet i, with a send time  $t_{i_s}$ and receive time $t_{i_r}$, we concatenate the packet delay $(t_{i_r} - t_{i_s})$, the ratio of packet reception interval $(t_{i+1_r} - t_{i_r}) / (t_{i+1_s} - t_{i_s})$, and the packet loss rate to form the probing flow. Sending $M$ consecutive probe packets constitutes one probing action, which takes time $T_s$. We conduct $N$ probing actions within the probing period $T_P$, where $T_P=\sum\limits_{}^{N} T_{s}$. The values of $M$ and $N$ are chosen to satisfy the requirements of the network state's dynamic changes.

\section{METHODOLOGY}
\subsection{Addictive Congestion Status}
The proposed model involves two states for link status: congested and uncongested, with the path state determined by the collective states of all links that comprise it. If any link within a path is congested, the entire path is considered congested. Conversely, a path is none-congested only if all its links are in a none-congested state. We introduce $x_i$ and $y_j$ to represent the boolean states of the $i$-th link and the $j$-th path, respectively. Specifically, $xi=1$ indicates that the i-th link is congested, and $xi=0$ indicates that the link is not congested, while we have
\begin{equation}
    y_{j}\triangleq \left\{ 
    \begin{array}{ll}
        0, &none-congested~\mathrm{for}~\sum_{l_{i} \preceq p_{j}}x_{i} = 0; \\
        1, &congested~\mathrm{for}~\sum_{l_{i} \preceq p_{j}}x_{i}\ge 1, \\
    \end{array} \right.
\end{equation}
where $l_{i} \preceq p_{j}$ means the $i$-th link routed by the $j$-th path.

\begin{figure}[t]
    \centering
    \begin{minipage}[c]{0.28\textwidth}
        \centering
    \includegraphics[width=\textwidth]{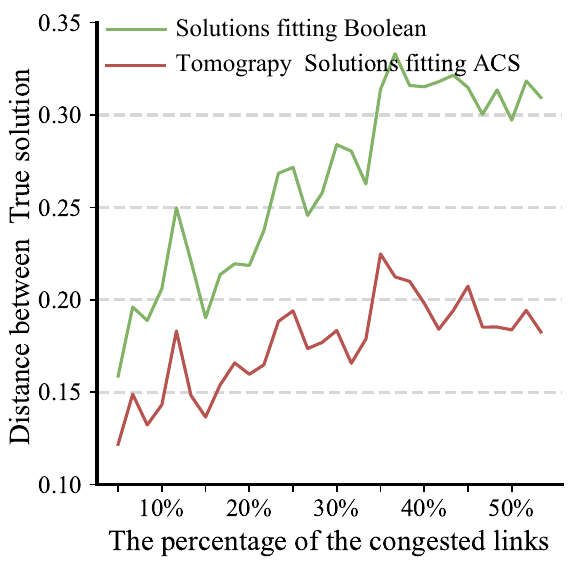}
    \subcaption{}
    \label{fig: distance}
    \end{minipage}
    \begin{minipage}[c]{0.19\textwidth}
            \centering
        \includegraphics[width=\textwidth]{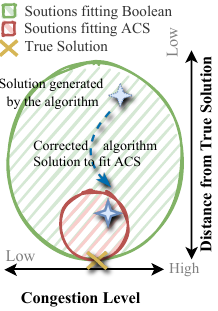}
        \subcaption{}
        \label{fig: solution-space}
    \end{minipage}
    \label{fig: sub}
    \caption{A brief comparison between the boolean and ACS based tomography. (a) Average distance between the tomography result and the true one. (b) Differences between the boolean and the ACS based solution spaces.}
\end{figure}

For the sake of the qualitative analysis, the \textbf{A}ddictive \textbf{C}ongestion \textbf{S}tatus (ACS) is introduced and comprises three types: none-congested, single-congested, and multiple-congested. Additionally, the $\mathcal{A}_{j}^+$ directly corresponds to the number of congested links in the j-th path and is mathematically formulated as follows:
\begin{equation}
   \mathcal{A}_{j}^+=\sum_{l_{i} \preceq p_{j}}x_{i},
    \label{eq: additive}
\end{equation}
while we also derive
\begin{equation}
\mathcal{A}_{j} \triangleq \left\{ 
\begin{array}{ll}
0, &none-congested~\mathrm{for}~{\mathcal{A}_{j}^+ = 0}; \\
1, &single-congested~\mathrm{for}~{\mathcal{A}_{j}^+=1}; \\
2, &multiple-congested~\mathrm{for}~{\mathcal{A}_{j}^+ \ge 2}. \\
\end{array} \right.
\end{equation}
For the sake of discussion, we specifically mark $\mathcal{A}=(\mathcal{A}_{1}^+, \mathcal{A}_{2}^+, \mathcal{A}_{2}^+, \cdots)$ as ACS+ for all observed paths.

Let vectors $Y$ and $X$ represent the path and link states, respectively. Then, the Addictive Congestion Status ($\mathcal{A}$) and \textbf{B}oolean \textbf{C}ongestion \textbf{S}tatus ($\mathcal{B}$) are calculated as 
\begin{gather}
    \mathcal{A}^+ = R \cdot X ;
    \label{eq: linear-additive}
    \\
    \mathcal{B} = R \bigcup X,
    \label{eq: OR-dot}
\end{gather}
where $\cdot$ denotes the dot product and $\bigcup$ is an element-wise logical OR operation between the products of the corresponding elements of the two vectors, and $X=(x_1, x_2, x_3,\cdots)$ collects all unknown links' statuses.

Unlike the BCS information (congested or not congested), ACS utilizes more granular path congestion information to assist the existing network tomography algorithms in diagnosing congested links or inferring link performance. \figref{fig: solution-space} illustrates the proposed scheme, where the green and red areas represent the solution spaces satisfying the path congestion observation constraints and ACS constraints, respectively. It is evident why the red area is smaller than the green area, as satisfying ACS inherently satisfies BCS, but the reverse is not true. If the solution of a network tomography algorithm does not meet the ACS constraints, i.e., it lies outside the red circle, the algorithm must adjust its solution until it meets ACS. When path congestion observations are accurate, the true solution (the actual congestion status of the link) will be at the bottom of the red circle because it satisfies both ACS and BCS  constraints and has the smallest distance to itself, which is zero. For this case, distance is defined by
\begin{equation}
    d(x, x_{i}) =  1 - \frac{\sum_{i=1}^{n} 1\left(x_{i}=x_{i}^{\prime}\right)}{n},
    \label{eq: define-distance}
\end{equation}
where  $x_{i}$ and  $x_{i}^{\prime}$  are the true i-th link state and the algorithm's estimated i-th link state, respectively, and n is the number of links in the topology. Through simulations conducted under different topologies and congestion probabilities (see \ref{Topolgy Setting} for experimental setup). As depicted in \figref{fig: distance}, the average distance between all solutions within the ACS-constrained solution space and the true solution is smaller than the average distance between all solutions that satisfy only the BCS constraint but not the ACS constraint and the true solution. This infers that solutions meeting the ACS constraint are closer to the true solution than those satisfying only the BCS constraint.

\begin{proposition}
	$\forall \mathbf{a}, \exists \mathbf{b}^{\prime}$, \text{ such that } 
 
 $d\left(\mathbf{a}, \mathbf{b}^{\prime}\right)<d(\mathbf{a}, \mathbf{b}) \Longrightarrow \mathcal{M}\left(\mathbf{a}, \mathbf{b}^{\prime}\right) \geq \mathcal{M}(\mathbf{a}, \mathbf{b})$
\end{proposition}
 $\mathcal{M}(\mathbf{a}, \mathbf{b})$ represents the performance metrics (such as recall, precision, and F1-score).

\begin{Proof}
If $b'$ has a smaller distance to $a$ than $b$: $d(a, b') < d(a, b)$.

This implies:
$\sum_{i=1}^{n} 1\left(a_{i}=b_{i}^{\prime}\right) > \sum_{i=1}^{n} 1\left(a_{i}=b_{i}\right).$

In order to understand how this impacts the performance metrics, consider the implications on True Positives (TP), False Positives (FP), and False Negatives (FN): $b'$ has more elements where $a_{i} = b_{i}^{\prime} = 1$; therefore: $
\mathrm{TP}_{\mathrm{b}^{\prime}} \geq \mathrm{TP}_{\mathrm{b}}.$

Similarly, for FP and FN: $
\mathrm{FP}_{\mathrm{b}^{\prime}} \leq \mathrm{FP}_{\mathrm{b}}, \quad \mathrm{FN}_{\mathrm{b}^{\prime}} \leq \mathrm{FN}_{\mathrm{b}}.$

Given these relationships:

\textbf{Recall} increases or remains the same as TP increases and FN decreases.
$
\text{Recall}_{\mathrm{b}^{\prime}} = \frac{\mathrm{TP}_{\mathrm{b}^{\prime}}}{\mathrm{TP}_{\mathrm{b}^{\prime}} + \mathrm{FN}_{\mathrm{b}^{\prime}}} = \frac{1}{1 + \frac{\mathrm{FN}_{\mathrm{b}^{\prime}}}{\mathrm{TP}_{\mathrm{b}^{\prime}}}}  \geq  \frac{1}{1 + \frac{\mathrm{FN}_{\mathrm{b}}}{\mathrm{TP}_{\mathrm{b}}}} = \frac{\mathrm{TP}_{\mathrm{b}}}{\mathrm{TP}_{\mathrm{b}} + \mathrm{FN}_{\mathrm{b}}} = \text{Recall}_{\mathrm{b}}.
$

\textbf{Precision} increases or remains the same as TP increases and FP decreases.
$
\text{Precision}_{\mathrm{b}^{\prime}} = \frac{\mathrm{TP}_{\mathrm{b}^{\prime}}}{\mathrm{TP}_{\mathrm{b}^{\prime}} + \mathrm{FP}_{\mathrm{b}^{\prime}}} = \frac{1}{1 + \frac{\mathrm{FP}_{\mathrm{b}^{\prime}}}{\mathrm{TP}_{\mathrm{b}^{\prime}}}} \geq \frac{1}{1 + \frac{\mathrm{FP}_{\mathrm{b}}}{\mathrm{TP}_{\mathrm{b}}}} = \frac{\mathrm{TP}_{\mathrm{b}}}{\mathrm{TP}_{\mathrm{b}} + \mathrm{FP}_{\mathrm{b}}} = \text{Precision}_{\mathrm{b}}.
$

\textbf{F\(\beta\)-score} increases or remains the same since it is the harmonic mean of precision and recall, both of which have increased or remained constant.
$
F_{\beta}\text{-score}_{\mathrm{b}^{\prime}} = \frac{(1+\beta^{2}) \cdot \text{Precision}_{\mathrm{b}^{\prime}} \cdot \text{Recall}_{\mathrm{b}^{\prime}}}{\beta^{2} \cdot \text{Precision}_{\mathrm{b}^{\prime}} + \text{Recall}_{\mathrm{b}^{\prime}}} = \frac{1}{\frac{\beta^{2}}{(1+\beta^{2}) \cdot \text{Recall}_{\mathrm{b}^{\prime}}} + \frac{1}{(1+\beta^{2}) \cdot \text{Precision}_{\mathrm{b}^{\prime}}}} \geq \frac{1}{\frac{\beta^{2}}{(1+\beta^{2}) \cdot \text{Recall}_{\mathrm{b}}} + \frac{1}{(1+\beta^{2}) \cdot \text{Precision}_{\mathrm{b}}}} = \frac{(1+\beta^{2}) \cdot \text{Precision}_{\mathrm{b}} \cdot \text{Recall}_{\mathrm{b}}}{\beta^{2} \cdot \text{Precision}_{\mathrm{b}} + \text{Recall}_{\mathrm{b}}} = F_{\beta}\text{-score}_{\mathrm{b}}.
$

The above validation demonstrates that the algorithm performs better when the ACS constraint is included.
\end{Proof}

\subsection{Long Short-Term Memory Network}

The LSTM network, a specific type of the Recurrent Neural Network (RNNs), operates by passing information from previous time steps to the current time step. LSTM has demonstrated impressive performance in handling spatio-temporal sequence problems. Its cell structure is meticulously designed to overcome the vanishing or exploding gradients problem that RNNs face when processing long sequences.

\begin{figure}[t]
	\centering
	\includegraphics[width=\columnwidth]{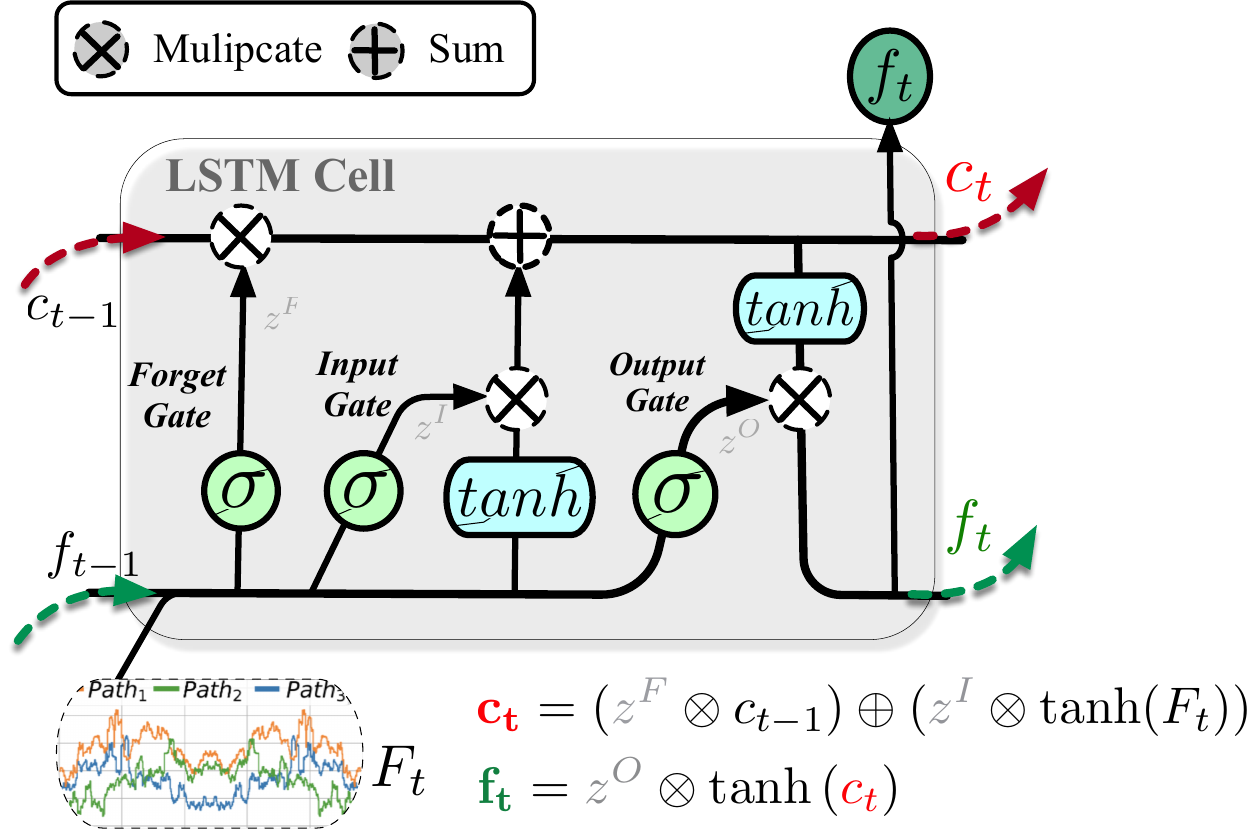}
	\caption{Illustration of the LSTM cell.}
	\label{fig: LSTM}
\end{figure}

The developed scheme considers the number of probing actions completed within the probing time $T_P$ as the number of time steps in the LSTM network and the amount of data collected during each probing action as the length of each time step. All observational data are normalized before being input into the model for training.

The LSTM network primarily comprises three stages:
\begin{enumerate}
    \item \textbf{Forget Gate}: Used for selectively forgetting the output from the previous time step, \(c_{t-1}\). Its function can be interpreted as "forgetting the unimportant". \(z^{F}\) acts as the forget gate control.
    \item \textbf{Input Gate}: Selectively remembers the input for the current time step, \(F_{t}\). Its function can be interpreted as "remembering the important". \(z^{I}\) acts as the memory gate control, participating in the update of the cell state to obtain the new state \(c_{t}\).
    \item \textbf{Output Gate}: Determines the output value \(f_{t}\) based on the cell state at the current time step.
\end{enumerate}

\begin{algorithm}[t]
    \SetKwInOut{Input}{Input}
    \SetKwInOut{Output}{Output}
    \SetKw{KwBreak}{\textbf{break}}

    \SetKwFunction{False}{False}
    \SetKwFunction{True}{True}

    \SetKwFunction{Clustering}{K-Means}
    \SetKwFunction{NT}{Network Tomography}

    \caption{Adversarial LSTM-based Addictive Congestion Status (ALACS)}
    \label{alg: p}
    
    \Input{ routing matrix $\mathbf{R}$, path set $\mathbf{P}$ \\ number of paths $P$, probing times $\mathcal{N}$}

    \BlankLine
    \tcc{Monitor Paths' Status}
    \BlankLine

    Infer the sets of Paths' Probing Information  $\mathbb{F} \leftarrow \emptyset$\;

    \For{t=1 \KwTo $\mathcal{N}$}{    
        \ForEach{$p_j \in \mathcal{P}$}{
            $F_j(t)$ $\leftarrow$ Probing and collecting the Information \;
        }
        $\mathbb{F}$ $\leftarrow$ ${\bm [}\mathbb{F}$, $\{F_j(t) | j=1, \cdots, P\} {\bm ]}$\;
    }
    
    \BlankLine
    \tcc{Calculate Paths' Congestion Status}
    \BlankLine

    Initialize the sets of Paths' Congesiton Status $\mathbb{S} \leftarrow \emptyset$\;
    
    \ForEach{$p_j \in \mathcal{P}$}{
        $\mathcal{A}_j$ $\leftarrow$ Obtain the \textit{Addictive Congestion Status} with model by $\{F_j(t) | t=1, \cdots, \mathcal{N}\}$\;
        
        $\mathbb{A}$ $\leftarrow$ ${\bm [}\mathbb{A}$, $A_{j} {\bm ]}$\;
    }

    \BlankLine
    \tcc{Infer the links' Performance}
    \BlankLine

    Initialize the sets of links' Performances by  $\mathbb{L} \leftarrow \emptyset$\;

    $\mathbb{L}$ $\leftarrow$ \NT{$\mathbb{F}$, $\mathbf{R}$, $\mathbb{A}$} \;

    \Output{$\mathbb{L}$}
\end{algorithm}

\subsection{Adversarial Autoencoder}
\begin{figure*}[t]
  \centering
  \begin{minipage}[b]{1\textwidth}
    \includegraphics[width=\textwidth]{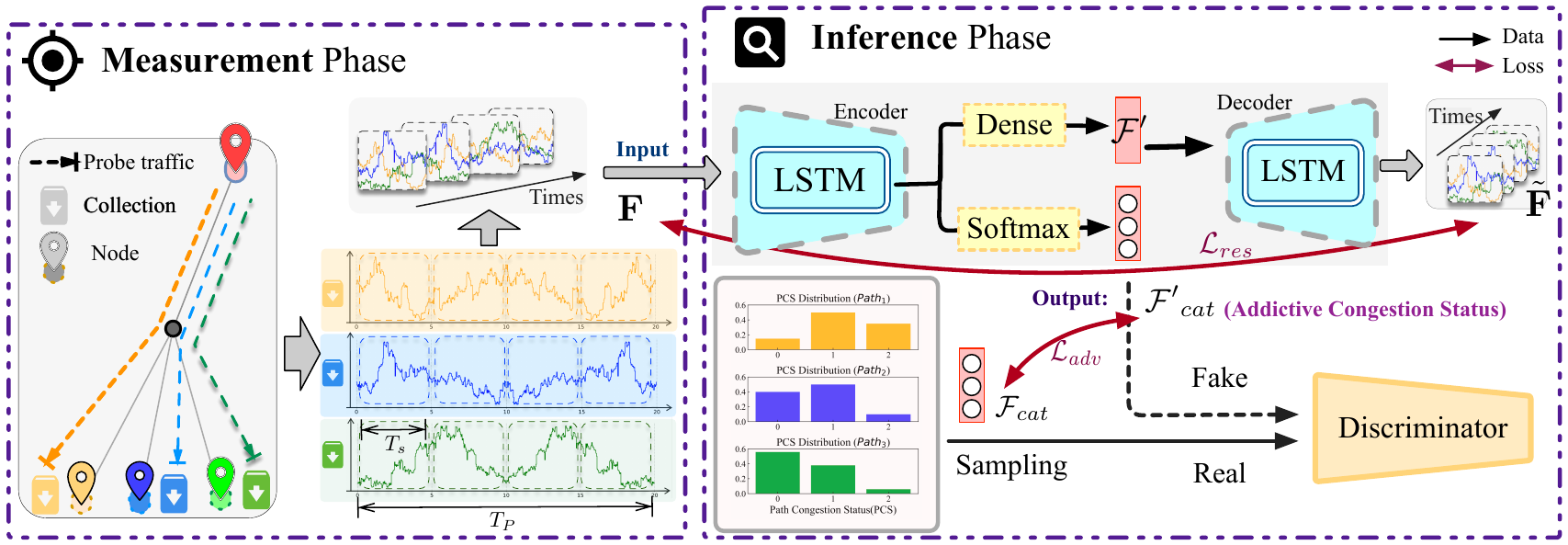}
    \caption{Overview of our proposed scheme. It has two phrases: one is illustrated on the left side, for the process of end-to-end measurements and data collection; the other is depicted on the right side, for the ACS path status identification/inference via AAE-LSTM based deep learning.}
    \label{fig: modeling}
  \end{minipage}

\end{figure*}
Adversarial Autoencoders (AAE)\cite{makhzani2015adversarial} are machine learning models that combine the features of Autoencoders (AE)\cite{autoencoder} and Generative Adversarial Networks (GANs) \cite{gans}. By integrating these two models, AAEs effectively learn the target data distribution. Autoencoders typically comprise an encoder and a decoder to learn an effective data representation (or features). The encoder compresses high-dimensional input data into a low-dimensional representation while the decoder reconstructs the original input data from this low-dimensional representation\cite{li2023ae-comprehensive}. The autoencoder aims to minimize the difference between the input and its reconstruction. Accordingly, GANs comprise a generator and a discriminator. The generator aims to produce fake data that is as realistic as possible, while the discriminator aims to distinguish between real and fake data effectively\cite{fei2023-gan}. Through this adversarial process, the generator eventually creates high-quality fake data.

Based on the data obtained from the detection flow, $\mathbf{F}$, the feature equation of the adversarial autoencoder based on LSTM is defined as: $f(\mathbf{F}) = \tilde{\mathbf{F}}$. In the autoencoder, the encoder and decoder are mirrors of each other, i.e., $\tilde{\mathbf{F}} = D(E(\mathbf{F}))$. Considering that the congestion characteristics of each time slot often correlate with the congestion characteristics before and after it, LSTM is employed as a part of the encoder and decoder. This strategy effectively encodes input data with temporal dependencies into the latent space, where $\mathcal{F}'$ and $\mathcal{F}_{cat}'$ are compressed representations of $\mathbf{F}$ from the latent space. We incorporate the GAN into the autoencoder to regulate the latent space's data distribution. This ensures that the distribution of the encoder's encoded results conforms to the distribution of Addictive Congestion Status information.

The training process of the proposed method is divided into the following two parts.

\subsubsection{Reconstruction Process} The autoencoder is trained by minimizing the difference between the input and the reconstructed output (e.g., using reconstruction error). Multiple LSTM layers are stacked in the encoder portion, where LSTM learns the temporal dependencies in the sequence through its internal state (hidden state and cell state), thereby converting the input sequence data into a fixed-size latent representation\cite{xin2023signal}. Then, the decoder decodes this latent representation back into the original data space.  Besides, the compressed vector from the encoder passes through the dense and softmax layers, generating continuous vectors $\mathcal{F}'$ and probability distribution vectors $\mathcal{F}_{cat}'$, respectively. Although both vectors possess the characteristics of observational data, $\mathcal{F}'$ is used for subsequent data reconstruction, while $\mathcal{F}_{cat}'$ is utilized for classification and specifically for ACS recognition.

\subsubsection{Adversarial Process} The latent representation is input to the adversarial network, which aims to differentiate between the distribution of the latent representation and the target distribution. The encoder in the autoencoder acts as the generator, attempting to produce latent representations that match the target distribution\cite{liu2024adversarial}. At the same time, the adversarial network tries to distinguish whether these representations are from the target distribution. The adversarial process aims to make the distribution of $\mathcal{F}_{cat}'$ approach the true distribution of Addictive Congestion Status.
Both the reconstruction loss and adversarial loss are defined as follows:
$$
\left\{
\begin{aligned}
    \mathcal{L}_{D^{1}}^{adversarial} = &\mathcal{L}_{\mathcal{F}_{cat}\sim \mathcal{N}}^{bce}(\mathcal{D}^{1} (\mathcal{F}_{cat}),1)) + \\ &  \mathcal{L}_{\mathbf{\mathcal{F'}_{cat}}\sim q(\mathcal{F'}_{cat})}^{bce}(\mathcal{D}^{1}(\mathcal{F'}_{cat}),0)); \\
	\mathcal{L}_{{LSTM-AAE}}^{rectruction} = &\mathcal{L}_{\mathcal{F}\sim q(\mathcal{F})}^{bce}(\mathcal{D}^{A}(\mathcal{F}),1))+\mathcal{L}^{mse}(\tilde{\mathbf{F}},\mathbf{F}).
\end{aligned}
\right.
$$
These losses will be used to guide the training of our proposed AAE-LSTM neural network as depicted in \figref{fig: modeling}, as similar to the parameter estimation of conventional regression problems like curve fitting.

\section{Numerical Evaluations}

This section demonstrates the impact of different detection flow settings on the experimental results and then evaluates the algorithm's ACS identification from both qualitative and quantitative perspectives. Finally, model ablation experiments prove the indispensable roles of LSTM and Adversarial Autoencoders.

\subsection{Network Setups} \label{Topolgy Setting}

All experiments were conducted with real network topologies obtained from TopologyZoo\cite{TopologyZoo}, aiming to assess the accuracy and reliability of identifying ACS. Each network topology had a distinct structure, such as varying path lengths and numbers of paths \ref{TopologyMetrics}. The network event simulations were conducted with NS-3 and the duration of congestion in each simulation instance was set to 5 minutes. All experimental data and related code are available online\footnote{\href{https://gitee.com/Monickar/acs}{https://gitee.com/Monickar/acs}}.
\begin{table}[!ht]
    \centering
    \caption{Characteristics of four real networks used for numerical simulations.}
    \scalebox{0.75}{
    \begin{tabular}{ccccc}
    \toprule
    \toprule
        \diagbox{Metrics}{Network} & \textbf{CHINANET} & \textbf{AGIS} & \textbf{GEANT} & \textbf{ERNET} \\ \midrule
        \textbf{\#Paths} & 17 & 14 & 15 & 12 \\ 
        \textbf{\#Links} & 21 & 18 & 17 & 13 \\ 
        \textbf{Average.Hops} & 3.9 & 3.6 & 3.6 & 3.25 \\ 
        \textbf{Average.Weights} & 4.3 & 2.8 & 3.1 & 3 \\ \hline
    \end{tabular}
    }
    \label{TopologyMetrics}
    
\end{table}

During the experiments, it was assumed that each link in the topology has a certain congestion probability, which is independent. Less than 20\% of the available bandwidth was used to determine whether a link is congested, i.e., when the available bandwidth of a link is between 100\% and 80\%, the link's status is considered normal. The link is considered congested when the available bandwidth is less than 20\%. The experiments were conducted on two topology types: \textit{Homogeneous Topology}, where all links in the network have the same propagation delay and bandwidth, both set at 20ms and 20Mbps, respectively and \textit{Heterogeneous Topology}, where the properties of links in the network vary. The propagation delay and bandwidth values follow a uniform distribution between 20 and 25. Unless specified otherwise, each scenario in the experiments, comprising the detection flow settings and the network topology, is repeated 40 times.

\subsection{Evaluations of ACS}

\subsubsection{Optimizing Probing Flow}
Prolonged monitoring and excessive probing rates can negatively impact network performance, while short-duration and low-rate probes may not effectively capture relevant network characteristics. We configured the probing flows regarding \textit{probing duration ratio}, \textit{Probe Traffic Bandwidth}, and \textit{windows number}, aiming to find an optimal set of probing settings to balance accuracy and intrusiveness.

\begin{figure}[t]
	\centering
	\includegraphics[width=\columnwidth]{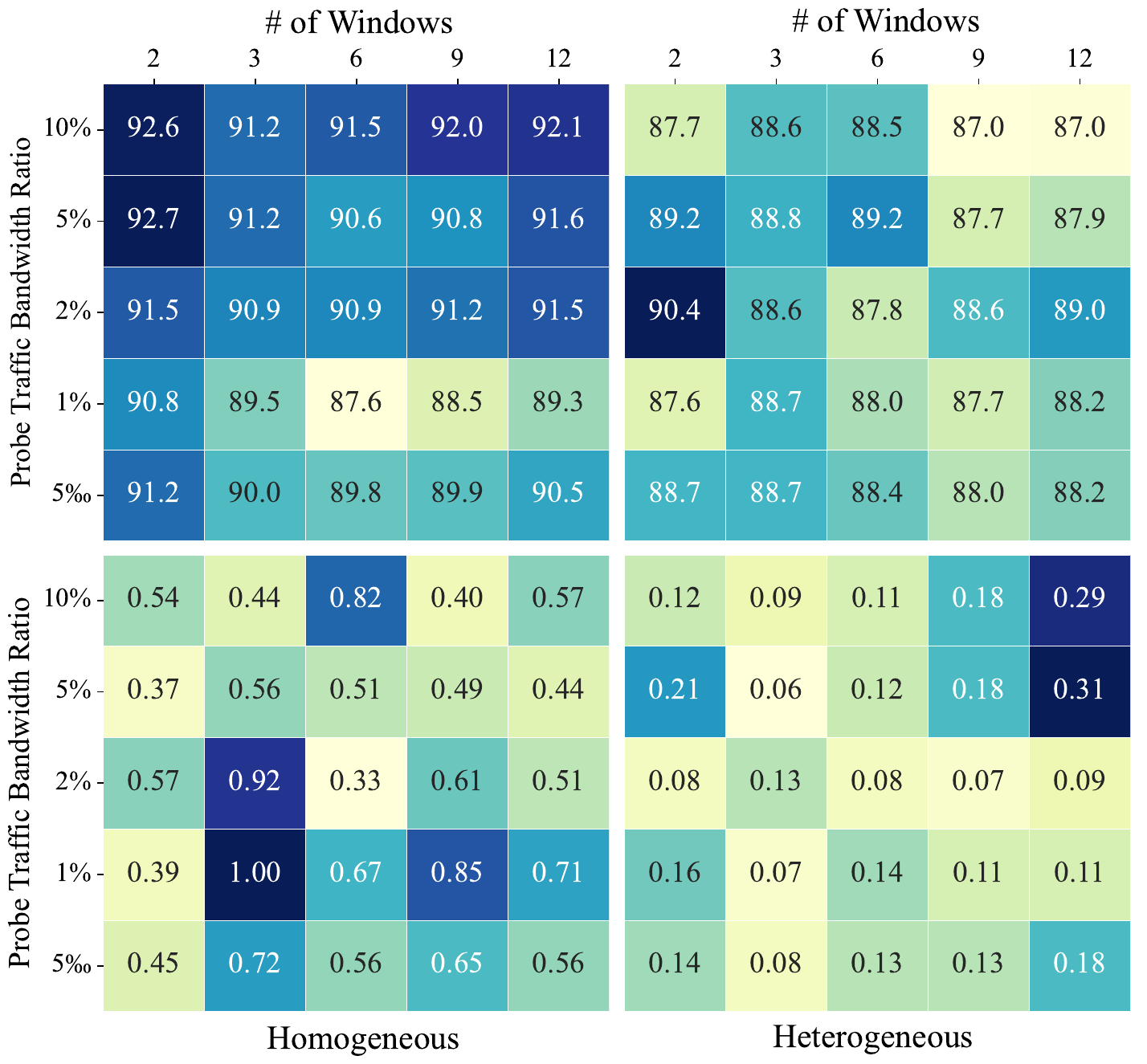}
	\caption{Accuracy of ACS identification under different probing conditions. Experiments were conducted in both homogeneous and heterogeneous network setups.}
	\label{fig: BestFlow}
\end{figure}

\figref{fig: BestFlow} reveals the impact of the \textit{Probe Traffic Bandwidth Ratio} (obtained by comparing the probing bandwidth to the end-to-end path bandwidth) and windows number on the accuracy and stability of Addictive Congestion Status identification. The area above indicates high accuracy (the darker the color, the higher the accuracy), and the area below represents the coefficient of accuracy variation (the lighter the color, the better the stability). It is evident that intense probing flows negatively affect background traffic by causing congestion, while too weak probes fail to capture congestion characteristics. Additionally, an excessive probe step size causes issues like data loss and gradient explosion in the LSTM model, while too small step sizes limit the LSTM's learning capability due to insufficient information, leading to poor training outcomes, low accuracy, and stability. Notably, compared to homogeneous topology, heterogeneous topology is more challenging in identifying Addictive Congestion Status.

\begin{figure}[t]
	\centering
	\includegraphics[width=\columnwidth]{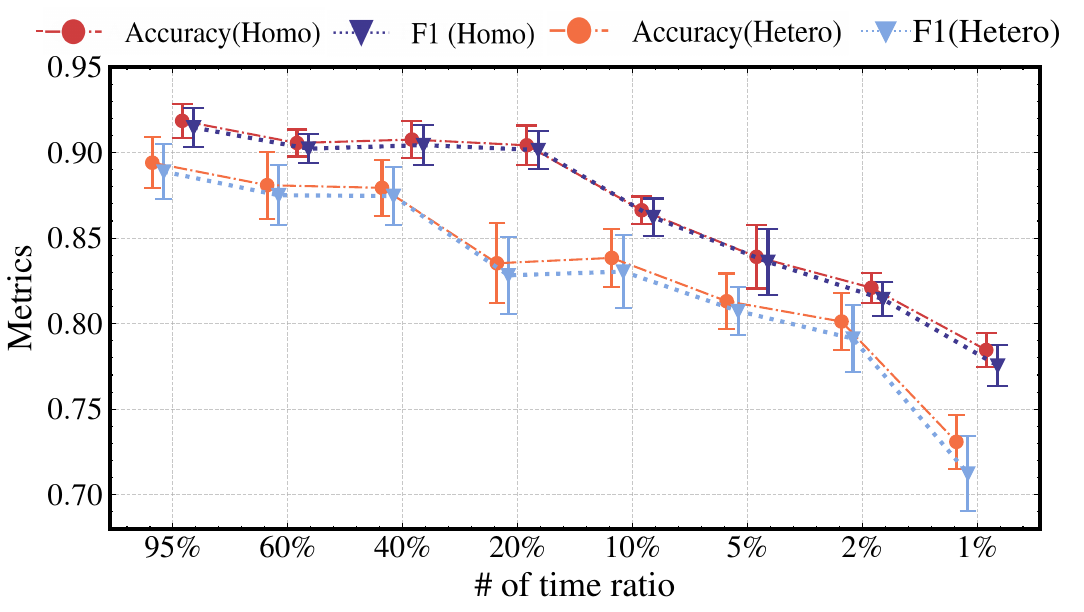}
	\caption{Accuracy and F1 scores of ACS based congestion diagnosis v.s. various  probing duration ratios.}
	\label{fig: FlowTime}
\end{figure}

\figref{fig: FlowTime} illustrates the effect of probing duration on the accuracy and stability of ACS using the accuracy and F1 score metrics to minimize the impact of sample imbalance on the results. Spherical markers represent accuracy, and triangular markers represent the F1 scores. The results reveal that as the probing duration decreases, accuracy and balance indices drop significantly when the observation time is less than 40\%. However, this is understandable, as longer probing times capture more congestion feature information.
Regarding the balance between intrusiveness, cost, and accuracy, the probing flow settings were 40\% probing duration, 1\% of average bandwidth for probe intensity, and a step size of 6.

\subsubsection{Categorized evaluation}
\begin{figure}[t]
    \centering
    \includegraphics[width=\columnwidth]{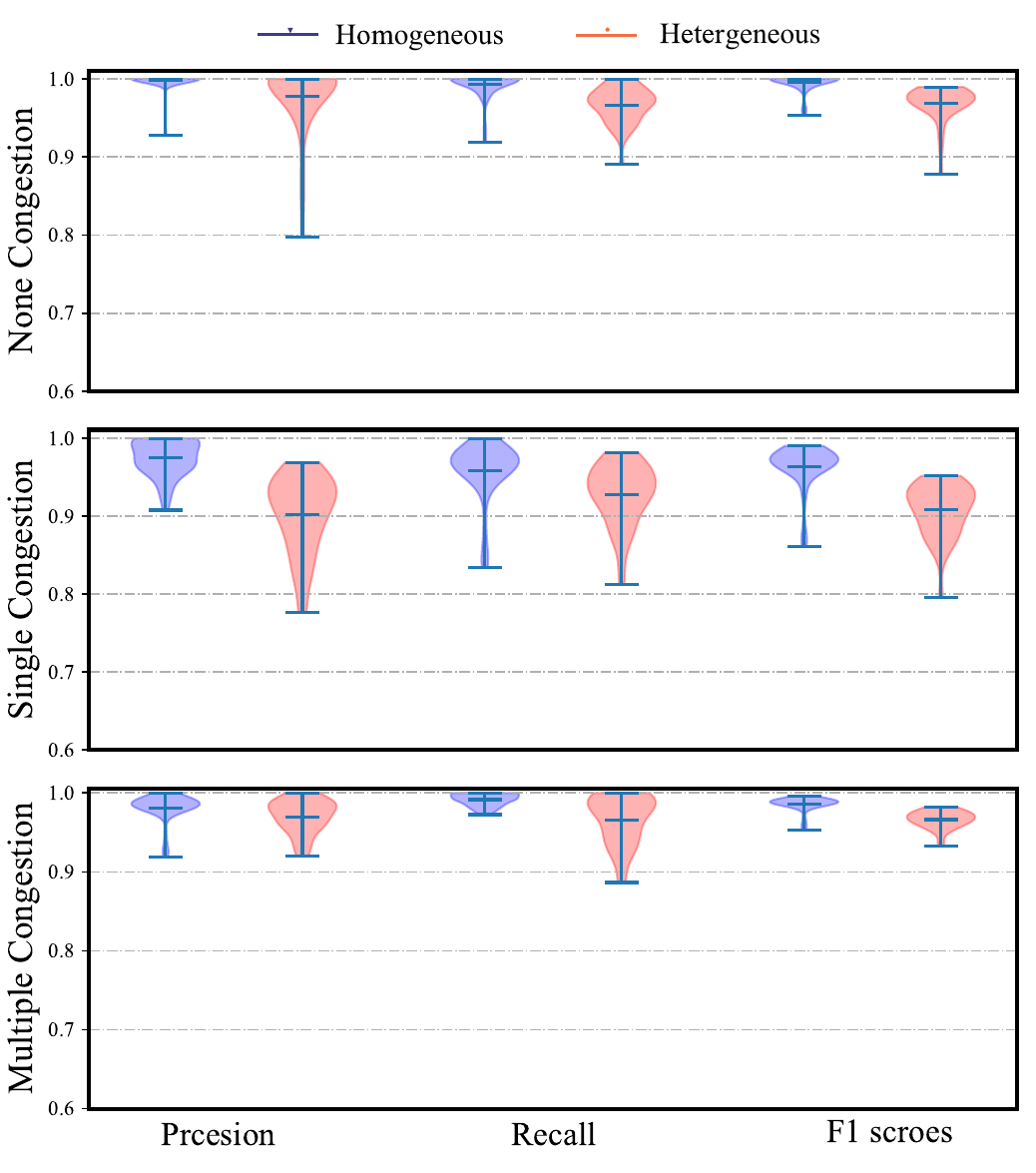}
    \caption{Distributions of precision, recall, and F1 scores across three categories of congestion: None, Single, and Multiple. The experiments are conducted in both homogeneous and heterogeneous network topologies.}
    \label{fig: dx}
\end{figure}

Congestion detection was considered a three-class problem, and we employed balanced sampling techniques for model training.
The proposed model was qualitatively evaluated in both homogeneous and heterogeneous network scenarios. The corresponding results are presented in \figref{fig: dx}, highlighting that our model exhibits high classification performance, with precision, recall, and average F1 scores all reaching or exceeding 0.95. However, compared to homogeneous topologies, heterogeneous topologies are more challenging for the model due to the diversity of link properties, resulting in lower performance across all metrics. It should be noted that the model demonstrates higher accuracy in identifying none-congested situations. However, distinguishing congestion further presents challenges for the model.

\subsubsection{Quantitative evaluation}
As the path lengths and the complexity of heterogeneous topologies increase, accurately identifying the number of congestions in a path becomes challenging. For instance, an algorithm might identify five congested links in a path where there are actually six. Hence, to demonstrate identification effectiveness, we use Relative Accuracy, with Absolute Accuracy and Relative Accuracy denoted as $\frac{\text{Number of Correct Classifications}}{N}$ and $\frac{\sum_{i=1}^{N}\left(1-\frac{\left|y_{i}-\hat{y}_{i}\right|}{\max (Y)-\min (Y)}\right)}{N}$, respectively.

\begin{figure}[t]
	\centering
	\includegraphics[width=\columnwidth]{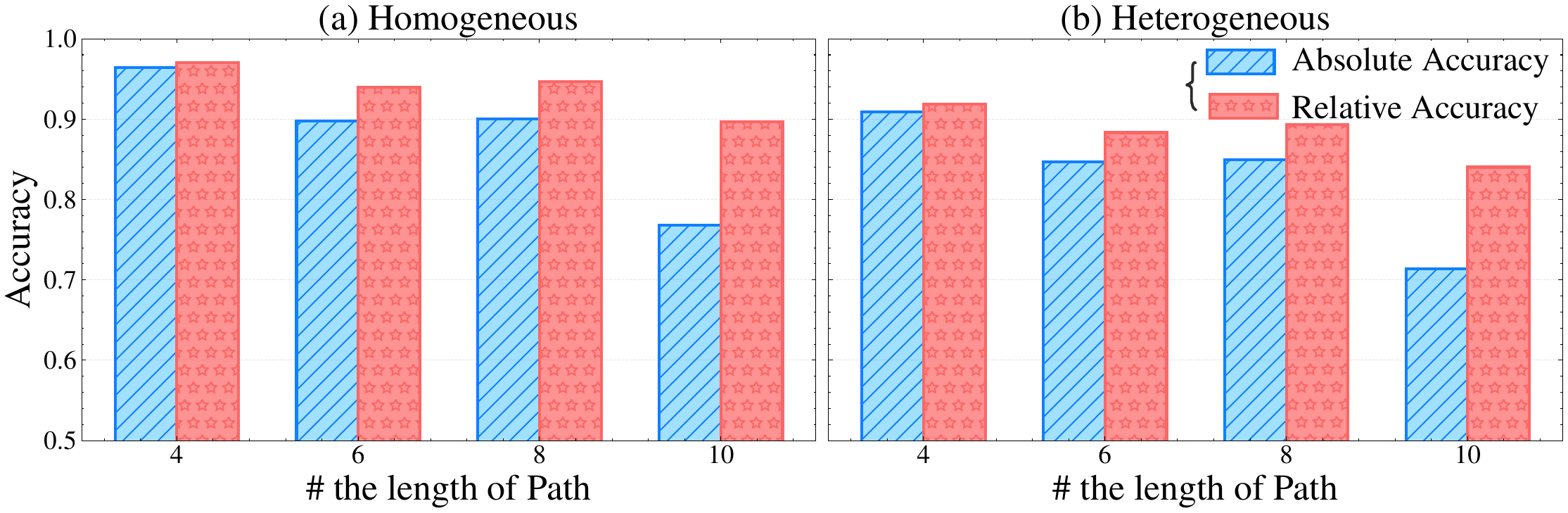}
	\caption{Estimation accuracy of absolute/relative path lengths from both homogeneous and heterogeneous setups of link prior congestion probability.}
	\label{fig: dl}
\end{figure}

\figref{fig: dl}  illustrates the relationship between the Absolute and Relative accuracy and path length in homogeneous and heterogeneous topologies. As path length increases, the model's absolute accuracy decreases, owing to the increased classification difficulty linked to the growing number of categories. However, the relative accuracy experiences a smaller decline, indicating that the model maintains a certain level of stability even as the data patterns become more complex.

\subsubsection{Ablation Studies}
To validate the efficacy of our framework combining AAE and LSTM networks, we conducted comparative experiments with four distinct models: LSTM only,  LSTM \& AAE, and AAE only (a fully connected layer substituted the LSTM). These experiments were meticulously designed with identical experimental conditions and sample data to isolate the impact of each model configuration on the accuracy of ACS recognition.

Our findings revealed a notable improvement in recognizing ACS when utilizing the combined strength of LSTM and AAE, as opposed to employing either LSTM or AAE in isolation. The LSTM model demonstrated its ability to capture temporal dependencies within the network data effectively. In contrast, the AAE model excelled in learning a comprehensive representation of the data distribution. However, the combination of the LSTM and AAE significantly enhanced the framework's capability to classify the ACS accurately, demonstrating the complementary strengths of these models in handling the complexity and variability inherent in network tomography data.

\begin{table}[t]
\centering
\caption{Ablation results from both homogeneous and heterogeneous setups of link prior congestion probability.}
\scalebox{0.75}{
    \setlength{\tabcolsep}{.7mm}
            \begin{tabular}{c c c |  c c c c c}
            \toprule
            \toprule
           \rowcolor{color3} &  AAE & LSTM & \textbf{Precision}  & \textbf{Recall} & \textbf{F1}-score & \textbf{ACC(\%)} & \textbf{R-Acc(\%)} \\
            \midrule
    \multirow{3}{*}{\rotatebox[origin=c]{90}{\textbf{Homo.}}} 
    
    &   & \checkmark & $0.939_{\pm 0.03}$  & $0.946_{\pm 0.03}$ & $0.936_{\pm 0.04}$ & $89.0_{\pm 5.3}$ & $93.0_{\pm 3.1}$\\
    &   \checkmark & & $0.927_{\pm 0.09}$  & $0.935_{\pm 0.07}$ & $0.926_{\pm 0.08}$ & $80.2_{\pm 9.9}$ & $89.4_{\pm 6.9}$\\
    &   \checkmark & \checkmark & \cellcolor[HTML]{C5F4F2} $\mathbf{0.984_{\pm 0.01}}$  &  \cellcolor[HTML]{C5F4F2} $\mathbf{0.981_{\pm 0.02}}$  & \cellcolor[HTML]{C5F4F2} $\mathbf{0.981_{\pm 0.01}}$  &  \cellcolor[HTML]{C5F4F2} $\mathbf{96.3_{\pm 2.4}}$  & \cellcolor[HTML]{C5F4F2} $\mathbf{97.8_{\pm 1.4}}$
    \\
    
            \midrule
    \multirow{3}{*}{\rotatebox[origin=c]{90}{\textbf{Hetero.}}} 
    
    &   & \checkmark & $0.829_{\pm 0.03}$  & $0.833_{\pm 0.02}$ & $0.816_{\pm 0.03}$ & $77.8_{\pm 3.4}$ & $83.5_{\pm 3.2}$\\
    &   \checkmark & & $0.736_{\pm 0.11}$  & $0.766_{\pm 0.07}$ & $0.712_{\pm 0.13}$ & $67.9_{\pm 8.9}$ & $78.5_{\pm 9.0}$\\
    &   \checkmark & \checkmark & \cellcolor[HTML]{C5F4F2} $\mathbf{0.949_{\pm 0.02}}$  &  \cellcolor[HTML]{C5F4F2} $\mathbf{0.953_{\pm 0.01}}$  & \cellcolor[HTML]{C5F4F2} $\mathbf{0.947_{\pm 0.02}}$  &  \cellcolor[HTML]{C5F4F2} $\mathbf{91.4_{\pm 3.0}}$  & \cellcolor[HTML]{C5F4F2} $\mathbf{94.1_{\pm 2.3}}$  \\

            \bottomrule
            
            \end{tabular}
    }

\end{table}

\subsection{Performance Improvement}
\begin{figure*}[t]
	\centering
	\includegraphics[width=\textwidth]{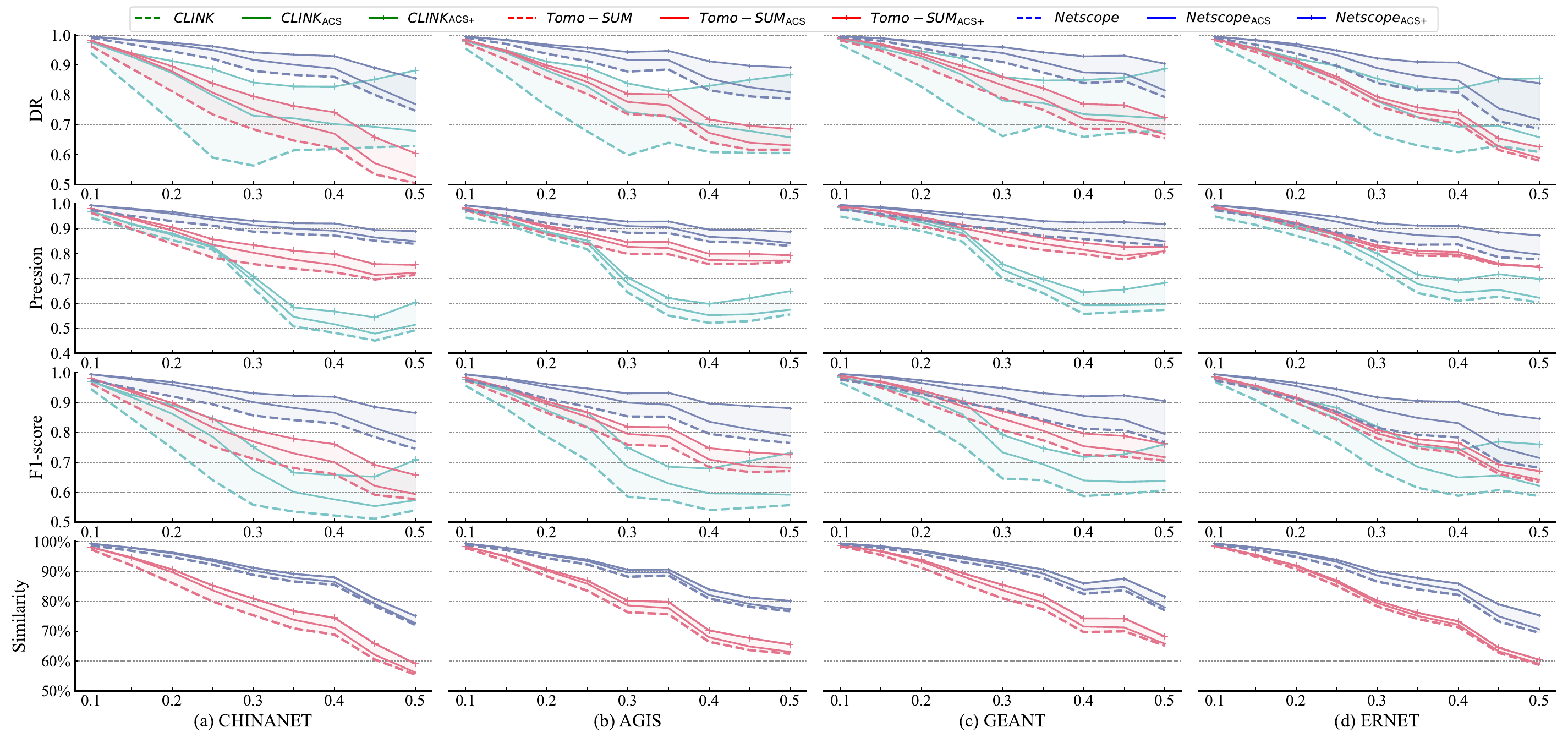}
	\caption{Performance comparisons under  various link congestion probabilities. The subscript ``ACS" and ``ACS+"  indicate that the diagnosis algorithm is fed with the ACS input of a path status $\mathcal{A}_{j} \in \{0, 1, 2\}$ and $\mathcal{A}_{j}^+ \in\{0, 1, 2, 3, ...\}$, respectively, while others employ the boolean path statuses.}
	\label{fig: improve_nt}
\end{figure*}

We verified the performance improvement of various algorithms constrained by the solution space using ACS in both congested link diagnosis and link performance inference (packet loss rate). Among the competitor algorithms, CLINK only has the function of congested link diagnosis. Sum-Tomo and Netscope first infer the link performance based on the observation information of end-to-end paths and then diagnose congestion based on the inferred link performance. For example, if a link's inferred packet loss rate exceeds 1\%, the link is considered congested.

In congested link diagnosis, we measure the diagnostic performance of the algorithms using recall, precision, and F1 score. The x-axis represents different scenarios of link congestion probabilities. The green, red, and blue plots represent the CLINK, Sum-Tomo, and Netscope algorithms, respectively. The dashed lines are the performance results without using ACS for the solution space constraints. The smooth solid lines represent the performance results of algorithms using qualitative ACS, and the solid lines with cross markers represent the performance results of algorithms using quantitative ACS.

\figref{fig: improve_nt} highlights that the performance of algorithms without using ACS for solution space constraint significantly declines as the congestion probability increases. Both the Sum-Tomo and Netscope aim to minimize the number of congested links, leading to potential miss-detections in high-congestion scenarios and consequently decreasing recall, resulting in a final decrease in the F1 score. In contrast, CLINK, as a greedy strategy based on the Maximum A Posteriori (MAP) algorithm that aims to find the most probable set of congested links according to the prior congestion probabilities of links, increases slightly in performance when the average congestion probability in the scenario exceeds 0.5. This is because its expected number of congested links increases in the congestion probability interval of [0.5-0.9] to satisfy the MAP criterion, thereby reducing miss-detections in high congestion scenarios and increasing recall.

Furthermore, as the link congestion probability increases, the performance improvement due to ACS also increases. This is because ACS helps algorithms address the pain points of uncertainty in network tomography by trimming the solution space, allowing the algorithms to avoid selecting sets with fewer congested links in scenarios with high congestion probabilities. Thus, DR increases with increasing link congestion probability. Notably, the precision of the algorithms (diagnosing congested links that are indeed congested) also increases because the solution space  reduces, ensuring the algorithms reduce their miss-detections while maintaining precision, thereby significantly enhancing the overall F1 score. 

Regarding link performance inference, this paper represents the estimation error by using the normalized root mean square error (NRMSE), which measures the similarity between two signals. For the true link performance $y[m]$ and inferred link performance $\hat{y}[m]$, their NRMSE is defined as follows: $$\mathrm{NRMSE}=\sqrt{\frac{\sum_{m=0}^{M-1} |y[m]-\hat{y}[m]|^{2}}{\sum_{m=0}^{M-1} |y[m]|^{2}}}$$
Netscope and Sum-Tomo present smaller estimation errors and a specific link performance in \figref{fig: improve_nt}. However, when ACS is inaccurate, its performance enhancement effect is understandably reduced, as incorrect ACS information introduces biases in trimming the solution space, potentially preventing the algorithms from finding solutions closer to the actual situation.

\section{CONCLUSION}

This paper has addressed the inherent challenges in network tomography, particularly the accurate identification of Addictive Congestion Status, by combining adversarial autoencoders with LSTM networks. The proposed framework leverages the spatio-temporal characteristics of network traffic, offering a robust solution for classifying and quantifying the Addictive Congestion Status. The proposed approach significantly enhances the network tomography's precision, mitigating the impact of the anomalous links on performance assessments while ensuring minimal invasiveness.

Experimental results demonstrate the efficacy of our method in reducing false and missed congestion detections, thereby validating its contribution to improving network performance evaluations. By capturing the dynamic nature of network traffic, our approach is significantly better than traditional network tomography techniques.

\section*{Acknowledgments}
We thank our colleagues and mentors for their valuable guidance during this research. We also appreciate the support from the High-performance Computing Platform of BUPT for providing the resources necessary to complete this work.

\bibliographystyle{unsrt}  
\bibliography{references}  

\vfill\pagebreak

\end{document}